\documentclass[pre,twocolumn,amssymb,noshowpacs,superscriptaddress]{revtex4}

\usepackage{graphicx}

\begin{document}

\title{Order 1 autoregressive process of finite length}

\author{C\u{a}lin Vamo\c{s}}
\affiliation{ ``T. Popoviciu'' Institute of Numerical Analysis, Romanian
   Academy, P.O. Box 68, 400110 Cluj-Napoca, Romania}
\author{\c{S}tefan M. \c{S}oltuz}
\affiliation{ ``T. Popoviciu'' Institute of Numerical Analysis, Romanian
   Academy, P.O. Box 68, 400110 Cluj-Napoca, Romania}
\affiliation{Departamento Matematicas, Universidad de los Andes,
   Carrera 1 No. 18A-10, Bogota, Columbia}
\author{Maria Cr\u{a}ciun}
\affiliation{ ``T. Popoviciu'' Institute of Numerical Analysis, Romanian
   Academy, P.O. Box 68, 400110 Cluj-Napoca, Romania}

\begin{abstract}
The stochastic processes of finite length defined by recurrence relations
request additional relations specifying the first terms of the process
analogously to the initial conditions for the differential equations.
As a general rule, in time series theory one analyzes only stochastic processes
of infinite length which need no such initial conditions and their properties
are less difficult to be determined. In this paper we compare the properties
of the order 1 autoregressive processes of finite and infinite length and we
prove that the time series length has an important influence mainly if the
serial correlation is significant. These different properties can manifest
themselves as transient effects produced when a time series is numerically
generated. We show that for an order 1 autoregressive process the transient
behavior can be avoided if the first term is a Gaussian random variable with
standard deviation equal to that of the theoretical infinite process and not
to that of the white noise innovation.

\end{abstract}

\maketitle

\section{\label{sec:level1}Introduction}

The purpose of the statistical processing of time series is to find a
stochastic model which can generate time series having the same statistical
properties as the observed series. One of the most common statistical model
is the autoregressive moving average (ARMA) process having many applications
in economy, finance, signal theory, geophysics, etc.\ \cite{box1976}. Lately,
especially the applications of the autoregressive (AR) models have become
more frequent in physics as well. For example they are used in the analysis
of the wind speed fluctuations \cite{hall2007}, radar signals \cite{gao2006},
climatic phenomena variability \cite{mar2004}, electroencephalographic
activity \cite{liley2003}, heart interbeat time series \cite{guzman2003},
the daily temperature fluctuations \cite{kiraly2002}, X-ray emmision from
the active galactic nuclei \cite{timmer2000}, sunspots variability
\cite{palus1999}, etc. As a result of their simple mathematical properties
and their direct physical interpretation, the realizations of AR(1)
processes have been used as artificial series to analyze some numerical
algorithms for monotonic trend removal \cite{vamos2007}, some surrogate
data test for nonlinearity \cite{kugiu2002} or for renormalization group
analysis \cite{blender2001}.

A stochastic process $\{X_{t},t=0,\pm1,\pm2,...\}$ is called autoregressive
process of order $p$, denoted AR$(p)$, if $\{X_{t}\}$ is stationary and for
any $t$ we have
\begin{equation}
X_{t}-\phi_{1}X_{t-1}-...-\phi_{p}X_{t-p}=Z_{t}~, \label{defARp}
\end{equation}
where $\{Z_{t}\}$ is a Gaussian white noise with zero mean and variance
$\sigma^{2}$, and $\phi_{1},\phi_{2},...,\phi_{p}$ are real constants. The
above relation may be rewritten in the compact form $\phi(B)X_{t}=Z_{t}$ where
\begin{equation}
\phi(z)=1-\phi_{1}z-...-\phi_{p}z^{p} \label{defpoly}%
\end{equation}
and $B$ is the shift operator, $BX_{t}=X_{t-1}$. Equation (\ref{defARp}) has
a unique solution if the polynomial (\ref{defpoly}) does not vanish for
$|z|=1$ \cite{brock1991}. If in addition $\phi(z)\neq0$ for all $|z|<1$,
then the process is causal, i.e., there exists a sequence of constants
$\{\psi_{j}\}$ such that $\sum_{j=1}^{\infty}|\psi_{j}|<\infty$ and
\begin{equation}
X_{t}=\sum_{j=0}^{\infty}\psi_{j}Z_{t-j}~,\qquad t=0,\pm1,... \label{defcauz}
\end{equation}
Hence, a causal process is characterized by the property that the random
variable $X_{t}$ can be expressed only in terms of noises at previous
moments and at the same moment. The properties of AR$(p)$ processes have
been studied in detail and they are the basis of the linear stochastic theory
of time series (\cite{brock1991}, \cite{brock1996}, \cite{Hamilt1994}).

The stochastic process defined above is an idealized mathematical object which
has no direct correspondent in practice. Each time series obtained by
measurements is a sequence of real numbers $x_{t}$ and one makes the assumption
that each of those numbers is a realization of a random variable $X_{t}$. To
recover the statistical information regarding the stochastic process we need a
large number of time series generated in the same conditions, a situation
occurring rarely in practice. Usually we have only one single time series and
the recognition of the stochastic process that generated it is very difficult.
In addition the finite length (sometimes very short) of the data series render
the problem even harder. Therefore it is important to analyze a finite
stochastic process satisfying the same recurrence relation as the idealized
infinite stochastic process (\ref{defARp}).

In the following we analyze the properties of an AR$(1)$ stochastic process
of finite length allowing us to make detailed analytical and numerical
computations. We compare the results obtained with the ideal case of the
infinite stochastic process and we discuss the conclusions which can be
obtained from its power spectrum.

\section{\label{sec:level2}AR(1) process}

A stochastic AR(1) process is defined by the simplified form of the relation
(\ref{defARp})
\begin{equation}
X_{t}=\phi X_{t-1}+Z_{t}~. \label{defAR1}
\end{equation}
By definition an AR(1) process has an infinite length.
In this case the polynomial (\ref{defpoly}) has the form $\phi(z)=1-\phi z$
and then the causality condition reduces to $|\phi|<1$. Because the relation
(\ref{defAR1}) is so simple, we can analyze more directly the significance of
this condition. By successively applying (\ref{defAR1}) $\tau$ times we obtain
\begin{equation}
X_{t}=Z_{t}+\phi Z_{t-1}+...+\phi^{\tau-1}Z_{t-\tau+1}+\phi^{\tau}X_{t-\tau}~.
\label{cauzrecur}%
\end{equation}
Because the process $\{X_{t}\}$ is stationary, the random variables $X_{t}$ and
$X_{t-\tau}$ have the same norm and then the norm of the last term in
(\ref{cauzrecur}) is $\phi^{\tau}$ times the norm of the left term. So for any
positive number $\varepsilon<1$, there exists $\tau>\ln\varepsilon/\ln|\phi|>0$
such that the last term from the right side can be neglected
\begin{equation}
X_{t}=Z_{t}+\phi Z_{t-1}+...+\phi^{\tau-1}Z_{t-\tau+1}+\mathcal{O}%
(\varepsilon) \label{AR1cauz}%
\end{equation}
for $\varepsilon\rightarrow0$.
This is the definition relation of a causal process (\ref{defcauz}) where
$\psi_{j}=\phi^{j}$. From relation (\ref{AR1cauz}) one observes that the
influence of the noise reduces as the time moves away from $t$.

In the nonstationary case $|\phi|=1$, all the terms in (\ref{cauzrecur})
have unitary coefficients and for any delay $\tau$ none of them can be
neglected. Every random variable $X_{t}$ is an infinite sum of terms with the
same norm, hence its norm is infinite.

\subsection{\label{sec:level2A} Causal AR(1) process}

In the following we analyze the basic properties of the causal stationary
process AR$(1)$ using the relation (\ref{defAR1}). If we take the mean of this
relation we obtain $E\left\{  X_{t}\right\}  =0.$ Its square is
\[
X_{t}^{2}=\phi^{2}X_{t-1}^{2}+2\phi X_{t-1}Z_{t}+Z_{t}^{2}.
\]
In accordance with relation (\ref{AR1cauz}) the random variables $X_{t-1}$ and
$Z_{t}$ are independent, that is $E\left\{  X_{t-1}Z_{t}\right\}  =0$. If we
take the mean of the last relation we obtain a relation between the variance
of successive random variables
\[
\sigma_{t}^{2}=\phi^{2}\sigma_{t-1}^{2}+\sigma^{2}.
\]
Because the AR$\left(  1\right)  $ process is stationary, it follows that for
every $t$ we have the same value for the variance $\sigma_{t}\equiv\sigma_{s}$
and then
\begin{equation}
\sigma_{s}^{2}=\frac{\sigma^{2}}{1-\phi^{2}}. \label{sigsta}%
\end{equation}

In order to compute the autocovariance function we multiply (\ref{defAR1})
with $X_{t-\tau}$ and we take the mean. Because the mean of $X_{t}$ vanishes
and $E\left\{  X_{t-1}Z_{t}\right\}  =0$, we obtain
\[
\gamma\left(\tau\right)\equiv E\left\{  X_{t}X_{t-\tau}\right\}  =\phi
\gamma\left(  \tau-1\right)  .
\]
By applying successively this relation and taking into account that
$\gamma\left(  0\right)  =\sigma_{s}^{2}$, we have
\begin{equation}
\gamma\left(  \tau\right)  =\sigma_{s}^{2}\phi^{\tau}~. \label{gamsta}%
\end{equation}
The spectral density is the Fourier transform of the autocovariance function
\begin{equation}
f\left(  \nu\right)  =\frac{1}{2\pi}\sum_{\tau=-\infty}^{\infty}e^{-2\pi
i\tau\nu}\gamma\left(  \tau\right)  , \label{TFcont}%
\end{equation}
where the positive number $\nu$ is the frequency. By direct calculation with
$\gamma\left(\tau\right)$ given by (\ref{gamsta}) we obtain
\begin{equation}
f\left(\nu\right) =\frac{\sigma^{2}}{2\pi}\frac{1}
{1-2\phi\cos2\pi\nu+\phi^{2}}. \label{densspectr}
\end{equation}
Due to the periodicity of this function, we reduce its domain of definition
to the interval $\nu\in [0,0.5]$.

When $\phi\rightarrow1$, the standard deviation $\sigma_{s}$ given by
(\ref{sigsta}) and the covariance function $\gamma(\tau)$ given by
(\ref{gamsta}) become both of them infinite. The spectral density
(\ref{densspectr}) becomes infinite only for $\nu=0$.

\subsection{\label{sec:level2B} Acausal AR(1) process}

Let us analyze the basic properties for the stationary acausal AR$(1)$
process, that is when $|\phi|>1$ in (\ref{defAR1}).
The relation (\ref{AR1cauz}) is no more true because now $\phi^{\tau}$
increases when $\tau$ increases. If we divide (\ref{cauzrecur}) by
$\phi^{\tau}$, then
\[
\phi^{-\tau}X_{t}=\phi^{-\tau}Z_{t}+\phi^{-\tau+1}Z_{t-1}+...+\phi
^{-1}Z_{t-\tau+1}+X_{t-\tau}~.
\]
For every positive $\varepsilon<1$, there exists $\tau>-\ln\varepsilon/\ln
|\phi|>0$ such that the term from the left side can be neglected and then
\begin{equation}
X_{t-\tau}=-\phi^{-1}Z_{t-\tau+1}-...-\phi^{-\tau+1}Z_{t-1}-\phi^{-\tau}%
Z_{t}+\mathcal{O}(\varepsilon)
\label{AR1acauz}%
\end{equation}
for $\varepsilon\rightarrow0$.
Hence, the random variable at moment $t-\tau$ can be expressed in terms of the
noise values at the future moments. It means that the direction of this
process is reversed to that of the causal process (\ref{defcauz}).

The same conclusion can be drawn from (\ref{defAR1}) if we write it in the form
\[
X_{t-1}=\phi^{-1}X_{t}-\phi^{-1}Z_{t}~.
\]
In this way we obtain a stationary causal process AR$(1)$ because $|\phi
^{-1}|<1$, but in reverse temporal direction and with the variance of the
noise $\phi^{-1}\sigma$. It means that the formulas for the acausal AR(1)
process can be obtained from those for the causal one by replacing
$\phi$ with $\phi^{-1}$ and $\sigma$ with $\phi^{-1}\sigma$. We shall
derive these formulas directly as well in order to verify this assertion.

\begin{figure*}
\includegraphics{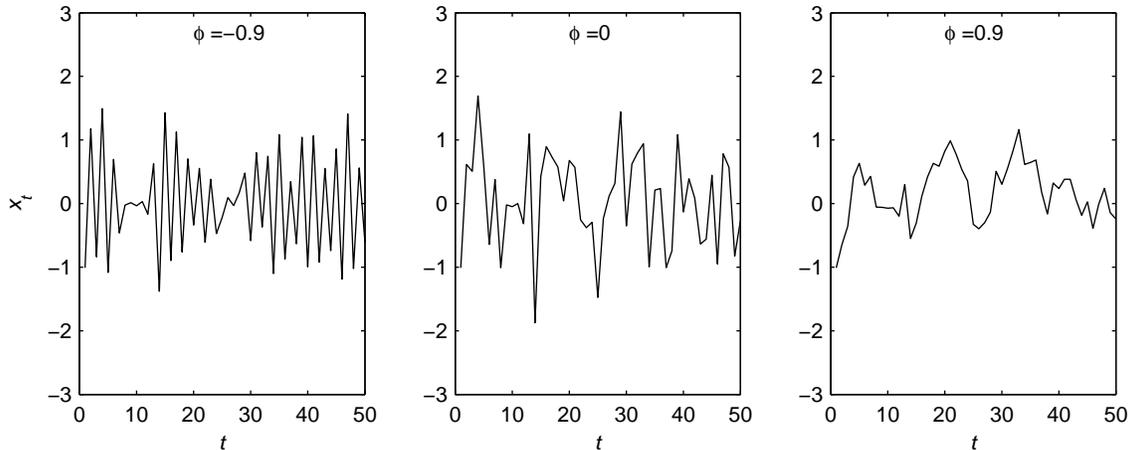}
\caption{\label{anticor} Realizations of an anticorrelated AR(1) process
($\phi=-0.9$), of an uncorrelated one ($\phi=0$), and of a correlated one
($\phi=0.9$).}
\end{figure*}

From (\ref{AR1acauz}) it follows that the random variables $X_{t-1}$ and
$Z_{t}$ are not independent, that is $E\left\{ X_{t-1}Z_{t}\right\} \neq0$.
In this case $E\left\{ X_{t}Z_{t}\right\} =0$, and then we write
(\ref{defAR1}) in the form
\[
\phi X_{t-1}=X_{t}-Z_{t}
\]
and we square it
\[
\phi^{2}X_{t-1}^{2}=X_{t}^{2}-2X_{t}Z_{t}+Z_{t}^{2}.
\]
If we take the mean instead of (\ref{sigsta}) we obtain
\[
\sigma_{s}^{2}=\frac{\sigma^{2}}{\phi^{2}-1}.
\]
In order to compute the autocovariance function we multiply (\ref{defAR1})
by $X_{t+\tau}$\ and instead of (\ref{gamsta}) we obtain
\[
\gamma\left(  \tau\right)  =\sigma_{s}^{2}\phi^{-\tau}~.
\]
Finally, relation (\ref{densspectr}) is the same.

\subsection{\label{sec:level2C} Anticorrelated AR(1) process}

Let us remark that the values of the parameter $\phi$ can be both positive and
negative. In order to qualitatively characterize the difference between the
two situations we use the fact that, if $\phi=0$, the AR(1) process is reduced
to a white noise with uncorrelated terms and a vanishing autocovariance
function for $\tau>0$. When $\phi>0$, from (\ref{defAR1}) it follows that the
fluctuations due to the white noise are superposed over the term
$|\phi|X_{t-1}$ which memorizes a part of the previous value of the time
series. Hence, the larger $\phi$ is, the closer from each other the successive
values of the time series are, and the fluctuations due to the white noise are
smaller. Therefore, in comparison with a realization of a white noise, for
$\phi>0$ the graphical representation of an AR(1) process is less fluctuant
and resembles to a deterministic trajectory disturbed by a random fluctuation
(see Fig.~\ref{anticor}). The autocovariance function (\ref{gamsta}) is positive
and tends to zero for $\tau\rightarrow\infty$.

If $\phi<0$, then the white noise is superposed over the term $-|\phi|X_{t-1}$
which has an opposite sign to the previous term of the time series.
Consequently the
white noise fluctuations are enhanced and the series values fluctuate with a
larger amplitude than the white noise, as shown in Fig.~\ref{anticor}. The
successive values of the autocovariance function (\ref{gamsta}) are of opposite
signs and the time series is called anticorrelated.

\section{\label{sec:level3}AR(1) power spectrum}

\begin{figure*}
\includegraphics{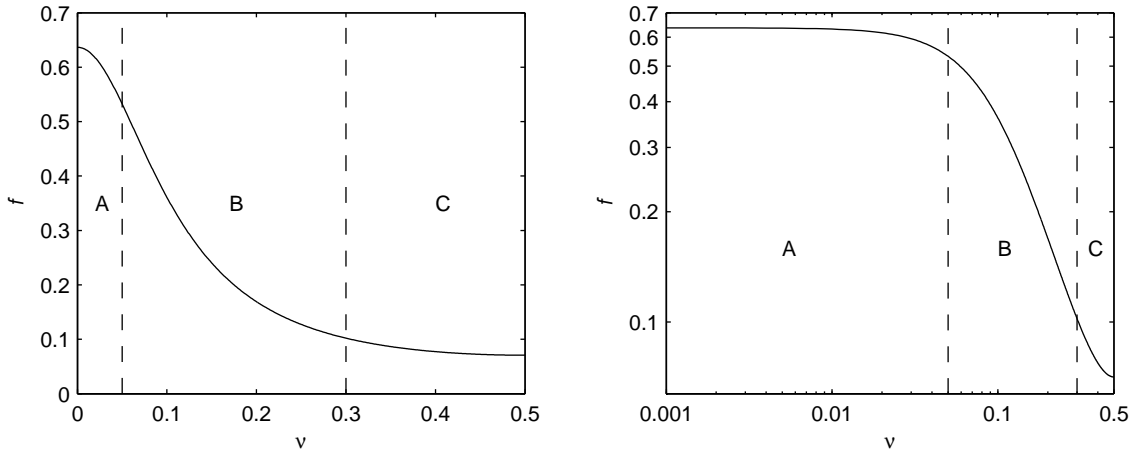}
\caption{\label{spectrufi05}The power spectrum of an AR(1) process for
$\phi=0.5$ and $\sigma=1$ on a linear scale and on a log-log one.}
\end{figure*}

In Fig.~\ref{spectrufi05} we have plotted the power spectrum
(\ref{densspectr}) for $\phi=0.5$ and $\sigma=1$ on a linear scale and on a
log-log one. The logarithmic coordinates strongly distort the shape of the
graphic because by taking the logarithm, the origin of the Ox axis is send
to $-\infty$ and any neighborhood of the origin is transformed into an infinite
length interval. We have separated the graphic into three regions (A, B, and C)
in order to evidence the deformations. For small frequencies
(region A) the spectral density is strongly stretched such that a plateau
appears with a value given by
\begin{equation}
f(0) =\frac{\sigma^{2}}{2\pi (1-\phi)^2}\ . \label{f0}
\end{equation}
From relation (\ref{densspectr}) one observes that the plateau corresponds to
the small values of $\nu$, when the variable term at the denominator can be
neglected in comparison with the constant term. Using the quadratic
approximation for cosine function we obtain the condition that the graph of
the AR$(1)$ power spectrum has a plateau
\begin{equation}
\nu \ll \frac{1-\phi}{2\pi\sqrt{\phi}}. \label{evalniu}%
\end{equation}
One remarks that if $\phi$ tends to 1, then the plateau appears at smaller
values of the frequency.

\begin{figure*}
\includegraphics{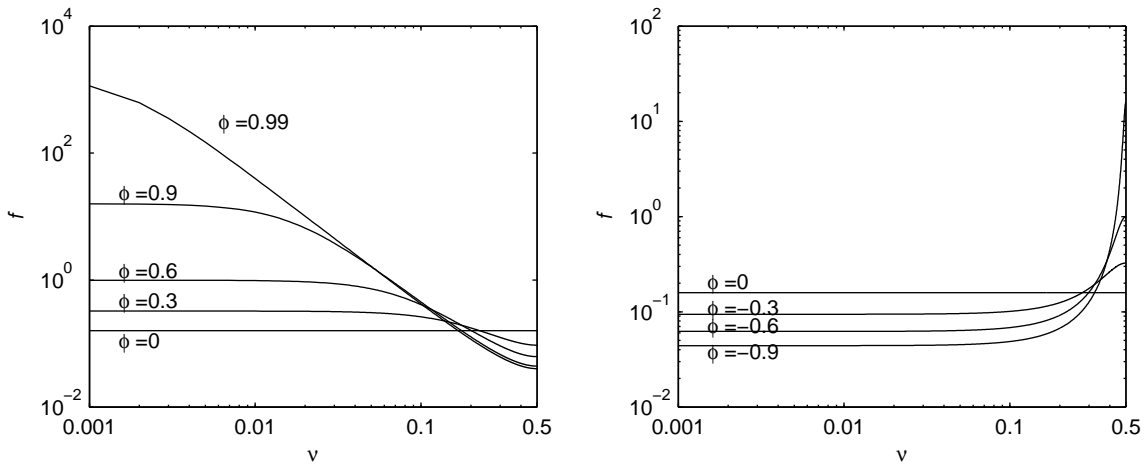}
\caption{\label{spectreAR1}Power spectra of an AR(1) process for $\sigma=1$
and different values of $\phi$.}
\end{figure*}

The region C of the spectral density for large frequencies is almost
parallel to the $Ox$ axis in linear coordinates. In logarithmic coordinates
it is strongly squeezed and acquires a significant slope. The region B of
median frequencies is not compressed so much, but its almost exponential
shape in linear coordinates becomes o curve with a considerable part having
a constant slope. The relative length of the three frequency regions depends
on the minimum value $\nu_{\min}$ of the frequency scale in the plot. If
$\nu_{\min}$ is not small enough, then the plateau may remain outside the
graphic.

As a global characteristic of the AR(1) power spectrum we introduce the
difference of its extreme values, quantity which we call the
\textit{spectrum amplitude} and is denoted by $A$. For $\nu\in [0,0.5]$ the
spectral density (\ref{densspectr}) is monotonic and its extreme values occur
at $\nu =0$ and $\nu =0.5$. Using (\ref{f0}) and
\begin{equation}
f(0.5) =\frac{\sigma^{2}}{2\pi (1+\phi)^2}\ , \label{f05}
\end{equation}
we have for the spectrum amplitude
\begin{equation}
A= |f(0)-f(0.5)| =\frac{4|\phi |\sigma^{2}}
    {\left(  1-\phi^{2}\right)  ^{2}}~. \label{amplit}%
\end{equation}
The extreme values of the power spectrum (\ref{f0}) and (\ref{f05}) become
equal for $\phi=0$ when the power spectrum is a line parallel with $Ox$ axis
and $A=0$. One can see that for $\phi\rightarrow1$, both $f(0)$ and $A$ tend
to infinity.

In Fig.~\ref{spectreAR1} we present the variation of the power spectrum with
respect to $\phi$. Although in linear coordinates the power spectrum
corresponding to $-\phi$ is the reflection of that corresponding to $\phi$
with respect to a line parallel to $Oy$ axis, in log-log scale they have very
different shapes. This difference occurs because, as shown in
Fig.~\ref{spectrufi05}, the small frequency region is stretched whereas that
of large frequency is compressed. In accordance with (\ref{f0}),
for $\phi>0$ the plateau
height $f(0)$ increases with $\phi$ and the extreme region for large
frequencies of the power spectrum $f(0.5)$ given by (\ref{f05}) has smaller
values for smaller $\phi$.

\begin{figure}
\includegraphics{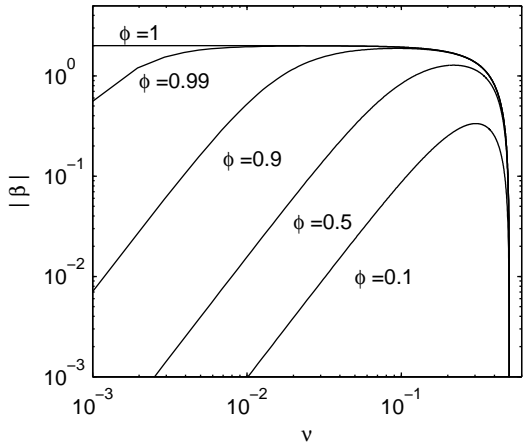}
\caption{\label{beta}The absolute value of the slope of the AR(1) power
spectrum for $\sigma=1$ and different values of $\phi$.}
\end{figure}

From Fig.~\ref{spectreAR1}a and Fig.~\ref{spectrufi05}b it results that the
AR(1) processes have some fractal features. For $\phi=0.9$ and especially
for $\phi=0.99$, a large region of the power spectrum is linear with a slope
near $-2$. Also for small values of $\phi$ (for example $\phi=0.5$ in
Fig.~\ref{spectrufi05}b) a significant region of the power
spectrum can be considered linear (fractal). In order
to compute the slope of the power spectrum for arbitrary $\phi$, we denote by
$\xi=\log x$ and $\eta=\log y$ the double-logarithmic coordinates such that
a function $y=f\left(  x\right)  $ is written in the new coordinates as
\[
\eta=\log_{10} f\left(  10^{\xi}\right)  .
\]
The slope of the log-log plot is
\[
\beta\left(  x\right)  =\frac{d\eta}{d\xi}=-x\frac{d}{dx}\left(  \ln f\left(
x\right)  \right)  .
\]
If we apply this relation to the function (\ref{densspectr}), we obtain the
slope of the AR(1) power spectrum
\begin{equation}
\beta(\nu ;\phi )=-\frac{4\pi\phi\nu\sin2\pi\nu}{1+\phi^{2}-2\phi\cos2\pi\nu}
        \ . \label{betaniu}
\end{equation}
In Fig.~\ref{beta} we have plotted the absolute value of this function in
log-log scale and we observe that for $\phi\gtrsim0.9$ there
exist values of $\beta$ near $-2$. We can verify this behavior substituting
$\phi =1$ in (\ref{betaniu})
\[
\beta(\nu ;1)=-2\pi\nu\cot\pi\nu\ .
\]
Obviously this case is artificial because
for $\phi=1$ we obtain the Brownian motion, not the AR(1) process.
Then we have $\lim_{\nu\rightarrow 0}\beta(\nu ;1)=-2$ which
corresponds to the plateau in Fig.~\ref{beta}. If
$\phi<1$, then $\beta (0;\phi )=0$ and in Fig.~\ref{beta} the curve is
decreasing for small frequencies.
For $\phi<0.9$ there is only one
maximum value for $\beta$ that corresponds to the center of the
\textquotedblright linear\textquotedblright\ (fractal) region of the power
spectrum.

\section{\label{sec:level4}Finite AR(1) process}

The time series appearing in practice have a finite length and usually they
are considered finite samples of an AR(1) process of
infinite length. The first terms of the sample are correlated with the
preceding terms of the realization which has not been recorded. But the first
terms of a numerically generated time series can not be related with
realizations of other preceding random variables. Therefore, a numerically
generated time series is never strictly a realization of a finite sample of an
ideal stationary stochastic process of infinite length. For instance, the index
$t$ in relations (\ref{defARp}) and (\ref{defAR1}) cannot be an arbitrary
integer. Because the relations defining the process are recursive, the first
terms must be defined by additional relations. As we shall show in the
following, the manner in which these additional relations are chosen can
essentially modify the properties of the stochastic process. We shall
call \emph{finite AR(1) process} a stochastic process of finite length
satisfying a recursive relation (\ref{defARp}).

Let $T$ denote the length of a finite AR(1) process, that is $t=1,2,...,T$.
The first term $X_{1}$ must be chosen independently and then
we obtain another process instead
of that studied in the previous section, denoted by $\{\widehat{X}_{t}\}$.
Because $\widehat{X}_{t}$ satisfies the relation (\ref{defAR1}) for $t>1$, if
we apply this relation successively we can express the terms of the stochastic
process as a finite sum
\begin{equation}
\widehat{X}_{t}=Z_{t}+\phi Z_{t-1}+....+\phi^{t-2}Z_{2}+\phi^{t-1}\widehat
{X}_{1}. \label{AR1finit}%
\end{equation}
In the following we consider only the causal AR$(1)$ processes, i.e.,
we suppose $|\phi|<1$. As shown in the previous section, the acausal process
is equivalent with a causal one generated in reverse order.

If $\widehat{X}_{1}$ is a Gaussian random variable with variance
$\widehat{\sigma}_{1}$ and zero mean, then from (\ref{AR1finit}) it follows
that $\widehat{X}_{t}$ is the sum of $t$ Gaussian random variables, hence
it has also a Gaussian distribution with variance
\[
\widehat{\sigma}_{t}^{2}=\sigma^{2}\left(  1+\phi^{2}+\phi^{4}+...+\phi
^{2(t-2)}\right)  +\widehat{\sigma}_{1}^{2}\phi^{2\left(  t-1\right)  }~.
\]
Applying the formula for the sum of a geometric series we have
\begin{equation}
\widehat{\sigma}_{t}^{2}=\sigma_{s}^{2}+\left(  \widehat{\sigma}_{1}%
^{2}-\sigma_{s}^{2}\right)  \phi^{2\left(  t-1\right)  }~, \label{sigsamp}%
\end{equation}
where we have used (\ref{sigsta}). The variance of the finite AR(1)
process has a constant term equal with the variance of the infinite AR(1)
process (\ref{sigsta}) and a variable term which tends asymptotically to zero
because $|\phi|<1$. In this case the finite AR(1) process is nonstationary
presenting transient effects, i.e., its variance approximates the theoretical
one $\widehat{\sigma}_{t}\simeq\sigma_{s}$ only after a time interval $t_{0}$
for which $\phi^{2t_{0}}$ can be neglected.

\subsection{\label{sec:level4A}Quasistationary finite AR(1) process}

For $\widehat{\sigma}_{1}=\sigma_{s}$ the variable term in (\ref{sigsamp})
vanishes and $\widehat{\sigma}_{t}=\sigma_{s}$ for all $t\leqslant T$.
Hence, if for a finite
AR(1) process we choose $\widehat{X}_{1}=(\sigma_{s}/\sigma)Z_{1}$, then
all the terms have the same variance. This choice is natural because it is
more reasonable to take the first term of the finite AR(1) process similar
to the stationary infinite AR(1) process and not to the white noise. Because
for $\widehat{\sigma}_{1}\neq\sigma_{s}$ the calculations become more
complicated, we deal only with the case $\widehat{\sigma}_{1}=\sigma_{s}$.

Let us show that for $\widehat{\sigma}_{1}=\sigma_{s}$ the properties of the
finite AR(1) process are very identical to those of a finite sample of a
stationary infinite AR(1) process. The autocovariance function
$\widehat{\gamma}(\tau ;t)=E\{\widehat{X}_t\widehat{X}_{t-\tau}\}$
can be calculated only if $|\tau|<T$ and $\tau<t\leqslant T+\tau$. Unlike the
autocovariance function (\ref{gamsta}) the quantity
$\widehat{\gamma}\left(  \tau;t\right)$ does not depend only on $\tau$ since
it exists only for certain values of $t$. Therefore $\{\widehat{X}_{t}\}$
is not a stationary stochastic process in a strict mathematical meaning.
However, when it exists, we can show that
$E\{\widehat{X}_{t}\widehat{X}_{t-\tau}\}=\phi E\{\widehat{X}_{t-1}\widehat
{X}_{t-\tau}\}$ proceeding in the same way as for (\ref{gamsta}). Then instead
of (\ref{gamsta}) we obtain
\begin{equation}
\widehat{\gamma}\left(  \tau;t\right)  =\phi^{\tau}\widehat{\sigma}_{t-\tau
}^{2}~. \label{gamfinit}%
\end{equation}
If we choose $\widehat{\sigma}_{1}=\sigma_{s}$, then
$\widehat{\sigma}_{t-\tau}=\sigma_{s}$ is constant and when
$\widehat{\gamma}\left(  \tau;t\right)$ exists it is identical to the
covariance function $\gamma\left(  \tau\right)$ in (\ref{gamsta}). Hence,
if we want to numerically model a stationary infinite AR(1) process,
then we have to use a finite AR(1) process $\{\widehat{X}_{t}\}$ with
$\widehat{\sigma}_{1}=\sigma_{s}$.

\subsection{\label{sec:level4B}Brownian motion}

Let us analyze now the finite AR(1) process satisfying (\ref{defAR1}) for
$\phi=1$. This is the well known Brownian motion. In this case
$\widehat{X}_{1}=Z_{1}$ and then (\ref{AR1finit}) becomes
\begin{equation}
\widehat{X}_{t}=Z_{t}+Z_{t-1}+...+Z_{2}+Z_{1}. \label{miscbrown}%
\end{equation}
Because $\widehat{X}_{t}$ is the sum of $t$ gaussian random variables, it
results that its variance is the sum of the variances of the terms of the sum
\begin{equation}
\widehat{\sigma}_{t}^{2}=t\sigma^{2}. \label{sigbrown}%
\end{equation}
If we write the relation (\ref{miscbrown}) in the form
\[
\widehat{X}_{t+\tau}=Z_{t+\tau}+...+Z_{t+1}+\widehat{X}_{t}~,
\]
multiply it with $\widehat{X}_{t}$ and take the mean, then we obtain
\begin{equation}
E\{\widehat{X}_{t+\tau}\widehat{X}_{t}\}=\widehat{\sigma}_{t}^{2}=t\sigma^{2}.
\label{covbrown}%
\end{equation}

The relation between the Brownian motion and the quasistationary finite AR(1)
process can be clarified if in (\ref{sigsamp}) we take
$\widehat{\sigma}_{1}=\sigma$ corresponding to the choice of the first term
for the Brownian motion $\widehat{X}_{1}=Z_{1}$
\begin{equation}
\widehat{\sigma}_{t}^{2}=\sigma_{s}^{2}\left(  1-\phi^{2t}\right)  ~.
\label{sigsampbrown}%
\end{equation}
Figure~\ref{beta} shows the variation of $\widehat{\sigma}_{t}^{2}$ for
different values of $\phi$. For a given $\phi$, at the beginning there is a
nonstationary transient period before the stationary  state of the AR(1)
process is reached. As $\phi$ tends to $1$, the transient region is expanded
and at the limit it becomes infinite, such that for $\phi=1$ an entirely
nonstationary process is obtained, i.e., the Brownian motion. So the Brownian
motion corresponds to the transient region extended to the infinity, whereas
the stationary infinite AR(1) process corresponds to the stationary part of
the graph. Therefore to obtain a quasistationary finite AR(1) process for
$\phi$ very close to $1$ the only possibility is to choose
$\widehat{\sigma}_{1}=\sigma_{s}$ completely eliminating the transient region.

\begin{figure}
\includegraphics{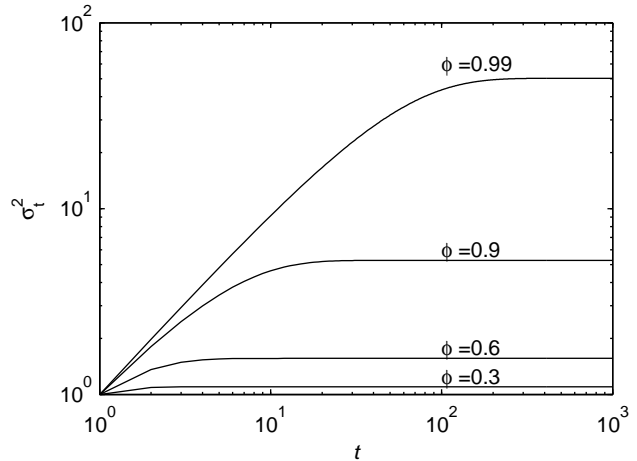}
\caption{\label{sigt}The standard deviation of a finite AR(1) process for
different values of $\phi$ when the first term coincides with the white noise
$\widehat{\sigma}_{1}=\sigma$. The transient region extends when $\phi$
takes greater values.}
\end{figure}

\section{\label{sec:level5}Periodogram}

For a finite AR(1) process the definition (\ref{TFcont}) can not be
applied. Therefore we use the periodogram which is defined as the square of
the absolute value of the discrete Fourier transform of the stochastic process
of finite length \cite{stoica1997}. Here we consider only a finite equidistant
set of frequencies for which we define the amplitude
\begin{equation}
A_{n}=\frac{1}{\sqrt{T}}\sum_{t=1}^{T}\widehat{X}_{t}\omega_{T}^{\left(
t-1\right)  \left(  n-1\right)  },\quad n=1,2,...,T, \label{FTdiscr}%
\end{equation}
where $\omega_{T}=\exp\left(  -2\pi i/T\right)  $. The corresponding terms of
the periodogram are
\begin{equation}
I_{n}=\left\vert A_{n}\right\vert ^{2}~. \label{peridgr}%
\end{equation}
Replacing (\ref{FTdiscr}) in (\ref{peridgr}) we have
\[
I_{n}=\frac{1}{T}\sum_{t=1}^{T}\sum_{s=1}^{T}\widehat{X}_{t}\omega
_{T}^{\left(  t-1\right)  \left(  n-1\right)  }\widehat{X}_{s}\omega
_{T}^{-\left(  s-1\right)  \left(  n-1\right)  }~.
\]
Because $\omega_{T}^{t-1}\omega_{T}^{-s+1}=\exp\left\{  -\frac{2\pi i}%
{T}\left(  t-s\right)  \right\}  $, we introduce a new variable $\tau=t-s$ and
then
\begin{equation}
I_{n}=\frac{1}{T}\sum_{t=1}^{T}\widehat{X}_{t}^{2}+\frac{1}{T}\sum_{\tau
=1}^{T-1}\left[  \omega_{T}^{\tau\left(  n-1\right)  }+\omega_{T}%
^{-\tau\left(  n-1\right)  }\right]  \sum_{t=1}^{T-\tau}\widehat{X}%
_{t}\widehat{X}_{t+\tau}. \label{peridgr2}%
\end{equation}
The terms $I_{n}$ of the periodogram are random variables and their mean
$E\left\{  I_{n}\right\}  $ tends to the spectral density $f(\nu )$ given by
(\ref{densspectr}) when $T\rightarrow\infty$ \cite{brock1991}. We shall show
this property only for the two particular cases considered in the previous
section for which the calculations are shorter.

\subsection{\label{sec:level5A}The periodogram of the quasistationary
        finite AR(1) process}

According to the analysis in the previous section, if
$\widehat{\sigma}_{1}=\sigma_{s}$, then
$E\{\widehat{X}_{t}^{2}\}=\sigma_{s}^{2}$ and
$E\{\widehat{X}_{t}\widehat{X}_{t+\tau}\}=\sigma_{s}^{2}\phi^{\tau}$
and the average of relation (\ref{peridgr2}) reads
\[
E\left\{  I_{n}\right\}  =\sigma_{s}^{2}+\frac{1}{T}\sigma_{s}^{2}\sum
_{\tau=1}^{T-1}\left(  T-\tau\right)  \phi^{\tau}\left[  \omega_{T}%
^{\tau\left(  n-1\right)  }+\omega_{T}^{-\tau\left(  n-1\right)  }\right]  .
\]
The sums in this formula are equal to
\begin{equation}
\sum_{\tau=1}^{T-1}\left(  T-\tau\right)  x^{\tau}=\frac{xT}{1-x}%
-\frac{x\left(  1-x^{T}\right)  }{\left(  1-x\right)  ^{2}} \label{sumatau}%
\end{equation}
where $x=\phi\omega_{T}^{\pm\left(  n-1\right)  }$. Since the first term is
proportional to $T$, it is larger than the second term,
therefore we write the periodogram mean as a sum of a dominant term and a
correction
\begin{equation}
E\left\{  I_{n}\right\}  =\frac{\sigma^{2}}{1-2\phi\cos\frac{2\pi}%
{T}(n-1)+\phi^{2}}(1+\rho_{n})~. \label{medprdg}%
\end{equation}
When the correction $\rho_{n}$ can be neglected, this formula is an
approximation of the spectral density of the stationary infinite AR(1) process
(\ref{densspectr})
\[
E\left\{  I_{n}\right\}  \simeq2\pi f\left(  \nu_{n}\right)  ,
\]
where $\nu_{n}=\left(  n-1\right)  /T$. Since (\ref{medprdg}) is a periodic
function, we shall consider for the index $n$ only the values
$n=1,2,...,[(T+1)/2]$, where $[\cdot]$ is the integer part function, so that
$\nu_{n}\in\lbrack0,0.5]$. Relation (\ref{medprdg})
can be written as well as
\begin{equation}
E\left\{  I_{n}\right\}  =\frac{\sigma^{2}}{(1-\phi)^{2}+4\phi\sin^{2}\pi
\nu_{n}}(1+\rho_{n})~. \label{medprdg2}%
\end{equation}

Now we shall analyze the dependence of the correction $\rho_{n}$ on the
parameters $n$, $T$ and $\phi$. Taking into account that in (\ref{sumatau})
$x^{T}=\phi^{T}$, from a direct computation it follows that $\rho_{n}$ can be
written as a product of two factors depending each of them only on two of the
three parameters
\begin{equation}
\rho_{n}=r(\phi,T)~\eta(\phi,\nu_{n})~, \label{ron}%
\end{equation}
where
\begin{equation}
r(\phi,T)=\frac{1}{T}\frac{2\phi(1-\phi^{T})}{1-\phi^{2}} \label{rcor}%
\end{equation}
and
\begin{equation}
\eta(\phi,\nu_{n})=\frac{2\phi-(1+\phi^{2})\cos2\pi\nu_{n}}{1+\phi^{2}%
-2\phi\cos2\pi\nu_{n}}~. \label{etan}%
\end{equation}

\begin{figure*}
\includegraphics{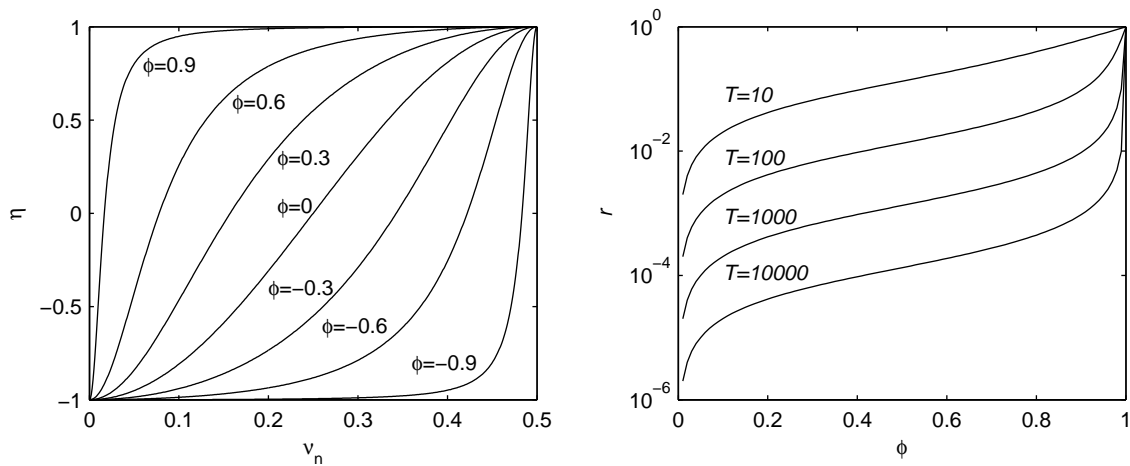}
\caption{\label{corectii} Correction of the periodogram of the quasistationary
finite AR(1) process in comparison with the power spectrum of the stationary
infinite AR(1) process.}
\end{figure*}

The derivative of the function $\eta$ with respect to $\nu_{n}$ indicates that
it is monotonic increasing for any fixed value $\phi\in(-1,1)$. For the extreme
values of $\nu_{n}$, the function $\eta$ is independent on the values of
$\phi$, i.e., $\eta(\phi,0)=-1$ and $\eta(\phi,0.5)=1$.
In Fig.~\ref{corectii} we have plotted $\eta$ with respect to $\nu_{n}$ for
different values of $\phi$. For $\phi=0$ the function reduces to a cosine
function $\eta(0,\nu_{n})=-\cos2\pi\nu_{n}$ which, as $|\phi|$ increases,
is distorted into a constant function, i.e.,
\begin{equation}
\lim_{\phi\rightarrow\pm1}\eta(\phi,\nu_{n})=\pm1~. \label{limeta}
\end{equation}
The absolute value of this factor is always smaller or equal to $1$.

The second factor in the correction (\ref{ron}) given by (\ref{rcor}) is
plotted in Fig.~\ref{corectii} for $\phi>0$ and a few values of $T$. It is
a monotonic increasing function with respect to $\phi$ and
\begin{equation}
\lim_{\phi\rightarrow1}r(\phi,T)=1~. \label{limr}
\end{equation}
From (\ref{limeta}) and (\ref{limr}) it follows that when
$\phi\rightarrow1$ the correction (\ref{ron}) is also approximately equal to
$1$, so that the approximation (\ref{medprdg}) of the AR(1) power spectrum is
wrong in the very dominant order. Reversely, for a given value of $\phi$,
the relation (\ref{rcor}) allows us to find the value of $T$ so that the
correction $r$ should have the desired value.

If $T$ is even, then these conclusions hold for $\phi<0$ too, since $r$ changes
only its sign $r(-\phi,T)=-r(\phi,T)$. But if $T$ is odd, then $\phi^{T}$
changes its sign and the function $r(\phi,T)$ acquires a different form. The
most important modification is that, for $\phi\rightarrow-1$ the factor $r$
becomes infinite. Thus, in this case, the periodogram of the quasistationary
finite AR(1) process is completely different from the power spectrum of the
stationary infinite AR(1) process.

\begin{figure}
\includegraphics{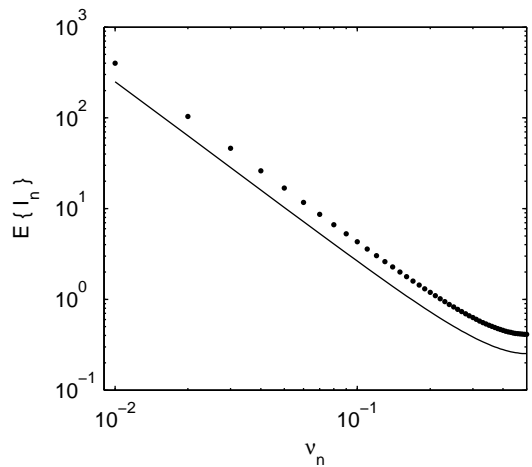}
\caption{\label{compsp} The periodogram of an finite AR(1) process with
$T=100$ and $\phi =0.99$ (point markers) in comparison with the power spectrum
of an infinite AR(1) process with the same $\phi$ (continuous line).}
\end{figure}

If the length $T$ is large enough such that the correction $\rho_{n}$ in
(\ref{medprdg2}) could be neglected, then the graphical representation of
the finite AR(1) power spectrum is identical to that discussed in
Section~\ref{sec:level3}. For example, when $T=1000$ and $\phi=0.9$ the
correction (\ref{ron}) is $\rho_n=0.0095\,\eta(\phi,\nu_{n})<0.0095$ and it
can be neglected. But for $T=1000$ and $\phi=0.99$ we have a much greater
correction $\rho_n=0.0995\,\eta(\phi,\nu_{n})$. In Fig.~\ref{corectii} one
can see that for $\phi$ near $1$, the factor $\eta$ is approximately $1$
excepting for very small frequencies. Therefore the correction $r$ is constant
for almost all frequencies and the periodogram (\ref{medprdg2}) differs from
the spectral density (\ref{densspectr}) by the constant factor $1+\rho_n$.
In log-log scale this constant factor shifts the periodogram parallel to the
AR(1) power spectrum (see Fig.~\ref{compsp}).

Because $\nu_1=0$, in a log-log scale the minimum frequency for a periodogram
of a finite AR(1) process is $\nu_2=1/T$. Then the condition that the
periodogram has a plateau can be obtained from (\ref{evalniu})
\begin{equation}
T \gg \frac{2\pi\sqrt{\phi}}{1-\phi}~.  \label{condT}
\end{equation}
For example, when $\phi =0.99$ we have $T \gg 600$ and in Fig.~\ref{compsp}
one can see that for $T=100$ the periodogram has no plateau.

\subsection{\label{sec:level5B}The periodogram of the Brownian motion}

The periodogram of the Brownian motion is obtained by replacing
(\ref{sigbrown}) and (\ref{covbrown}) in the formula resulting by averaging
(\ref{peridgr2})
\begin{eqnarray}
E\left\{  I_{n}\right\} &=&\frac{\sigma^{2}}{2}\left(  T+1\right)+
\frac{\sigma^{2}}{2T}\sum_{\tau=1}^{T-1}\left(  T-\tau\right)  \left(
T-\tau+1\right) \nonumber \\
&&\left(  \omega_{T}^{\tau\left(  n-1\right)  }+\omega
_{T}^{-\tau\left(  n-1\right)  }\right)  \label{EIn}.
\end{eqnarray}
After some calculations we obtain
\begin{equation}
E\left\{  I_{n}\right\}  =\frac{\sigma^{2}T}{8\sin^{2}\left(  \frac{\pi}%
{T}\left(  n-1\right)  \right)  }+\frac{\sigma^{2}\cos\left(  \frac{2\pi}%
{T}\left(  n-1\right)  \right)  }{2\sin^{2}\left(  \frac{\pi}{T}\left(
n-1\right)  \right)  }. \label{medInbrown}%
\end{equation}
The dominant term in (\ref{medInbrown}) is different from that in
(\ref{medprdg2}) for $\phi=1$ only by the factor $T/2.$ The relation
(\ref{medInbrown}) does not hold for $n=1$ because in the calculations we used
relations incompatible with that special value. If in (\ref{EIn}) we take
$n=1$, then
\begin{equation}
E\left\{  I_{1}\right\}  =\frac{\sigma^{2}}{3}\left(  T+1\right)  \left(
T^{2}-T+3\right)  . \label{EI1brown}%
\end{equation}

\begin{figure}
\includegraphics{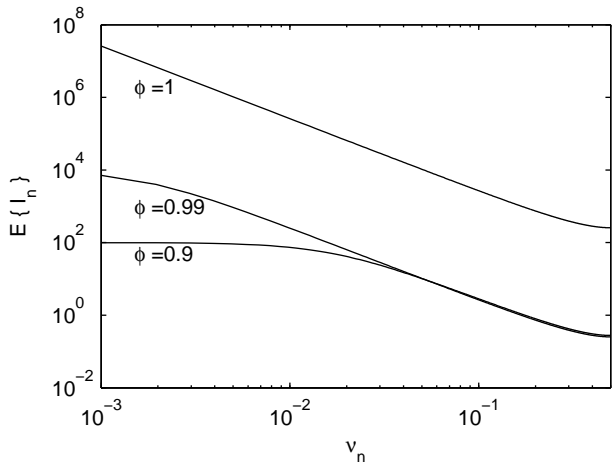}
\caption{\label{specbr}The periodogram of the Brownian motion in comparison
with the periodograms of the quasistationary finite AR(1) process with
$\phi=0.9$ and $0.99$.}
\end{figure}

In Fig.~\ref{specbr} we have plotted the dominant term of the periodogram
(\ref{medInbrown}) for $T=1000$. The most part of the graphic is a straight
line since, if in (\ref{medInbrown}) we consider $n\ll T$, then
$E\left\{I_{n}\right\}  \varpropto n^{-2}$. So in a log-log scale we obtain
a straight line with the slope $-2$ and since the small frequencies region
is strongly delated the most part of the graphic has this property.
Figure~\ref{specbr} shows that outside of the plateau, the periodogram of
the Brownian motion is parallel to the periodograms of quasistationary finite
AR(1) processes with $\phi$ close to $1$ and the distance between them equals
$\log(T/2)$ as shown above. This behavior is due to the fact that for
$1-\phi\ll1$ the value of the sine function at the denominator in
(\ref{medprdg2}) is dominant. Only for $\nu_{n}$ small enough the quantities
(\ref{medprdg2}) and (\ref{medInbrown}) are significantly different.

\section{\label{sec:level6}Conclusions}

Although the finite and infinite AR(1) processes are defined by the same
recurrence relation, their properties can be very different when the serial
correlation is large and the length of the finite AR(1) process is small.
This difference is minimized if the first term of the finite AR(1) process
is chosen such that its standard deviation should be equal to the standard
deviation of the infinite AR(1) process. However, even in this case the
periodogram of the finite AR(1) process for $\phi$ close to $\pm1$ can
significantly differ from the power spectrum of the infinite AR(1) process.
That is why to numerically generate a time series as a realization
of an AR(1) process we must take into account that the series length must be
larger than the threshold value (\ref{condT}).

Although the AR(1) process is the simplest stochastic process describing the
serial correlation only by means of a single parameter, it has remarkable
properties which make it very useful as a first step in time series modeling.
However, one has to take care that it is only one in the infinity of existing
stochastic models and it is possible that an autoregressive structure to be
incorrectly  assigned to a time series. For example, although the
autocovariance function (\ref{gamsta}) has an exponential decay, sometimes
one attempted to model a $1/f$ noise characterized by a power law decay
 with autoregressive processes \cite{kau1999}, \cite{coza2003}.
Therefore, it is necessary a thorough analysis of the AR(1) process properties
not only in its infinite idealized  form, but in the finite one as well.
If the time series is not long enough, it is very likely that essential
characteristics of the AR(1) process should be lost, as for example the
existence of the plateau at small frequencies, and misinterpretations can
occur.

\begin{acknowledgements}
This work was supported by MEdC under Grant 2-CEx06-11-96/19.09.2006.
\end{acknowledgements}

\end{document}